\documentclass[letterpaper,prl,preprint]{revtex4}
\usepackage{graphicx,bm}

\begin{document}

\title{Discreteness of populations enervates biodiversity in evolution}

\author{Yen-Chih Lin}

\author{Tzay-Ming Hong}
\email{ming@phys.nthu.edu.tw}

\author{Hsiu-Hau Lin}
\email{hsiuhau@phys.nthu.edu.tw}
\affiliation{Department of Physics, National Tsing Hua University, Hsinchu 30043, Taiwan}

\date{\today}

\begin{abstract}
Biodiversity widely observed in ecological systems is attributed to the dynamical balance among the competing species. The time-varying populations of the interacting species are often captured rather well by a set of deterministic replicator equations in the evolutionary game theory. However, intrinsic fluctuations arisen from the discreteness of populations lead to stochastic derivations from the smooth evolution trajectories. The role of
these fluctuations is shown to be critical at causing extinction and deteriorating
the biodiversity of ecosystem. We use children's rock-paper-scissors game to demonstrate how the intrinsic fluctuations arise from the discrete populations and why the biodiversity of the ecosystem decays exponentially, disregarding the detail parameters for competing mechanism and initial distributions. The dissipative trend in biodiversity can be analogized to the gradual erosion of kinetic energy of a moving particle due to air drag or fluid viscosity. The dissipation-fluctuation theorem in statistical physics seals the fate of these originally conserved quantities. This concept in physics can be generalized to scrutinize the errors that might be incurred in the ecological, biological, and quantitative economic modeling for which the ingredients are all discrete in number.
\end{abstract}
\maketitle

Biodiversity is commonly used to indicate the stability of an ecosystem\cite{May74,Pimm84,Jablonski08}. One of the central issues is to effectively promote the biodiversity while attracting more scientistsÕ attention from various fields\cite{McLaughlin02,Both06,Sala00, Reichenbach07, Loreau01}. The causes that threaten the biodiversity, for instance, climate change\cite{McLaughlin02,Both06}, over-harvesting, habitat destruction\cite{Sala00}, and population mobility\cite{Reichenbach07}, are well studied. Above those factors, Darwin's theory of natural selection plays a crucial role in catalysis\cite{Smith82,Hofbauer98,Nowak06,Nowak06a}. People are warned to reduce these effects in order to maintain and reserve the nature's biodiversity. Nevertheless, a naive reversed statement can be checked, namely, without any hazardous factors, would ecosystem be perfectly stable?

Here, we show the emergence of an intrinsic force originated from the fact that populations must be discredited. Furthermore, this force macroscopically jeopardizes the biodiversity of an ecosystem. However, populations in an ecosystem are discrete integers. Approximating these discrete populations by continuous variables inevitably introduces \textit{intrinsic fluctuations}, which turn the evolutionary dynamics stochastic in nature. When external noises are introduced, the stochastic process has been shown to be capable of causing mass extinction\cite{Bak93} in analogous to the avalanche in the sand piles. How important is this tiny difference between discrete and continuous variables when the population size is large? Do the intrinsic fluctuations simply introduce small irregularities or will they ever accumulate and cause a drastic impact on the biodiversity of the ecosystem?

To put the discussions on firm ground, we concentrate on the non-transitive rock-paper-scissors game\cite{Drossel01,Kerr02,Czaran02,Nowak04,West06,Szabo07}, known as a paradigm to illustrate the species diversity.
When three subpopulations interact in this non-transitive way, we expect that each species can invade another when its population is rare but becomes vulnerable to the other species when over populated.
The non-hierarchical competition\cite{Traulsen05,Frey08,Traulsen08,Frey09} gives rise to the endlessly spinning wheel of species chasing species and the biodiversity of the ecosystem reaches a stable dynamical balance.
This cyclic evolutionary dynamics has been found in plenty of ecosystems such as coral reef invertebrates\cite{Jackson75}, lizards in the inner Coast Range of California\cite{Sinervo96} and three strains of colicinogenic \textit{Escherichia coli}\cite{Kerr02,Kirkup04} in Petri dish.
Although the oscillatory solutions for the replicator equations capture the main features, inclusion of mobility\cite{Reichenbach07} or/and finite-population effects\cite{Traulsen08,Frey09} in the numerical simulations always jeopardizes the stable equilibrium and highlight the importance of stochasticity in the evolutionary dynamics.

To measure the effects due to the discreteness of the populations, we introduce a biodiversity indicator which is a direct product of all three subpopulations to specific powers and remains constant in the continuous replicative evolution.
By extensive numerical simulations, we record how the biodiversity indicator receives random corrections from the intrinsic fluctuations.
In addition to the irregular deviations at the short-time scale, it is truly remarkable that dissipative dynamics emerges as the stochastic processes accumulate and the biodiversity indicator thus decays exponentially. Slowly but surely, one species will
first become extinct after a half-time proportional to the population size\cite{Ifti03,Reichenbach06}. Our findings can be elegantly summarized in three steps: discreteness induces fluctuations, fluctuations spawn dissipations and dissipative dynamics leads to extinction.

The subtle connection between fluctuations and dissipations is best exemplified by a damped simple harmonic oscillator moving in a viscous liquid\cite{Reif08}.
The microscopic random bombardments from the thermal molecules coarse-grain into a macroscopic friction, causing the otherwise-conserved mechanical energy to dissipate exponentially. Although our analysis is based on the cyclic-competing ecosystem, the emergent dissipative dynamics from the intrinsic fluctuations can be readily applied to general biological and ecological systems. It can also be generalized to many other practices, such as the Turing\cite{Turning} model for biological patterns\cite{Pattern} or quantitative economic modelling on capital stock\cite{Stock} and reverse logistics\cite{logistics} where the basic ingredient is respectively the discrete pigment and monetary unit.

Here we start the exploration on the rock-paper-scissors game and investigate how the dissipative dynamics arises from the discreteness of the populations.
Consider three cyclically competing species $A$, $B$, $C$ with the stochastic interactions among them,
\begin{eqnarray}
A+B &\stackrel{k_{ab}}{\longrightarrow}& A+A,
\nonumber\\
B+C &\stackrel{k_{bc}}{\longrightarrow}& B+B,
\nonumber\\
C+A &\stackrel{k_{ca}}{\longrightarrow}& C+C,
\end{eqnarray}
where $k_{ab}, k_{bc}, k_{ca}$ are the relative probabilities for the cyclical replacements to occur.
For convenience, we choose the normalization $k_{ab}+k_{bc}+k_{ca}=1$.
Not that the stochastic processes preserve the total population $N=N_A+N_B+N_C$, where $N_A, N_B, N_C$ are the subpopulations for each species.
In each simulation time step, a pair of individuals is randomly chosen and evolves stochastically according to the cyclical competing processes.

To establish the connection between the stochastic and the replicative approaches, it is insightful to derive the effective replicator equations for the population ratios defined as $a(t) = N_A(t)/N$ and similarly for $b(t)$ and $c(t)$.
As long as the total population $N$ is large, we can approximate the time derivatives of the population ratios as $[\langle a(t+\tau) \rangle - \langle a(t) \rangle]/\tau$.
The detail derivations for the effective replicator equations scratched from the microscopic stochastic processes can be found in Supplementary Online Materials (SOM),
\begin{eqnarray}
\dot{a} &=& a(k_{ab} b - k_{ca} c) + \xi_a(t),
\nonumber\\
\dot{b} &=& b(k_{bc} c - k_{ab} a) + \xi_b(t),
\nonumber\\
\dot{c} &=& c(k_{ca} a - k_{bc} b) + \xi_c(t),
\end{eqnarray}
where $\xi_a(t), \xi_b(t), \xi_c(t)$ are intrinsic noises arisen from the discreteness of populations and only become insignificant in the limit of infinite population.
Even though these intrinsic noises are not correlated at different times, they are not completely independent.
Because $N_A+N_B+N_C$ is strictly conserved at each microscopic evolution step, a global constraint holds among these noises, $\xi_a(t) + \xi_b(t) +\xi_c(t) = 0$.

\begin{figure}
\centering
 \includegraphics[width=14cm]{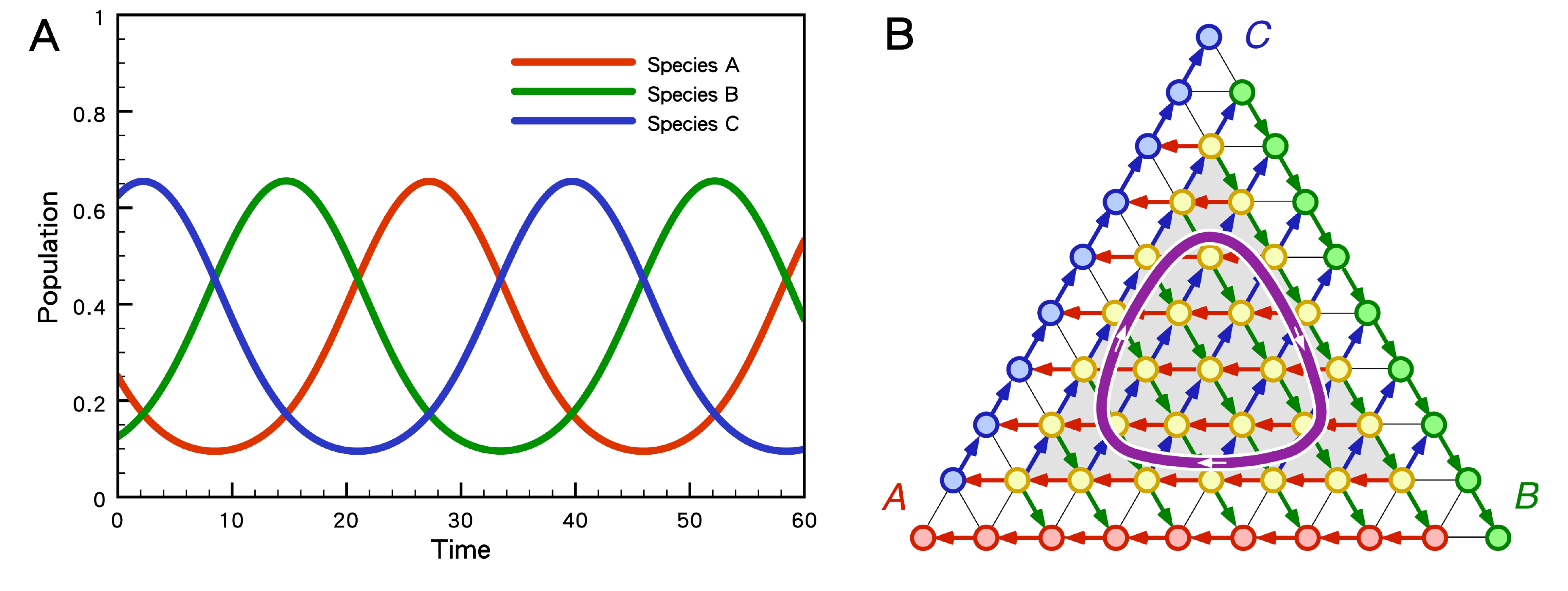}
\caption{Evolution for the cyclically competing species from replicator equations.
(A) The cyclic competition generates the delicate dynamical balances among the species and leads to the stable oscillatory populations for the competing species.
(B) The oscillatory population densities trace out a closed simplex contour (purple solid line) inside the triangle.
For finite population size $N=9$, the changes of the populations ratios are described by stochastic directional hopping on the underlying triangular lattice inside the simplex.}
\end{figure}

\begin{figure}
\centering
\includegraphics[width=10cm]{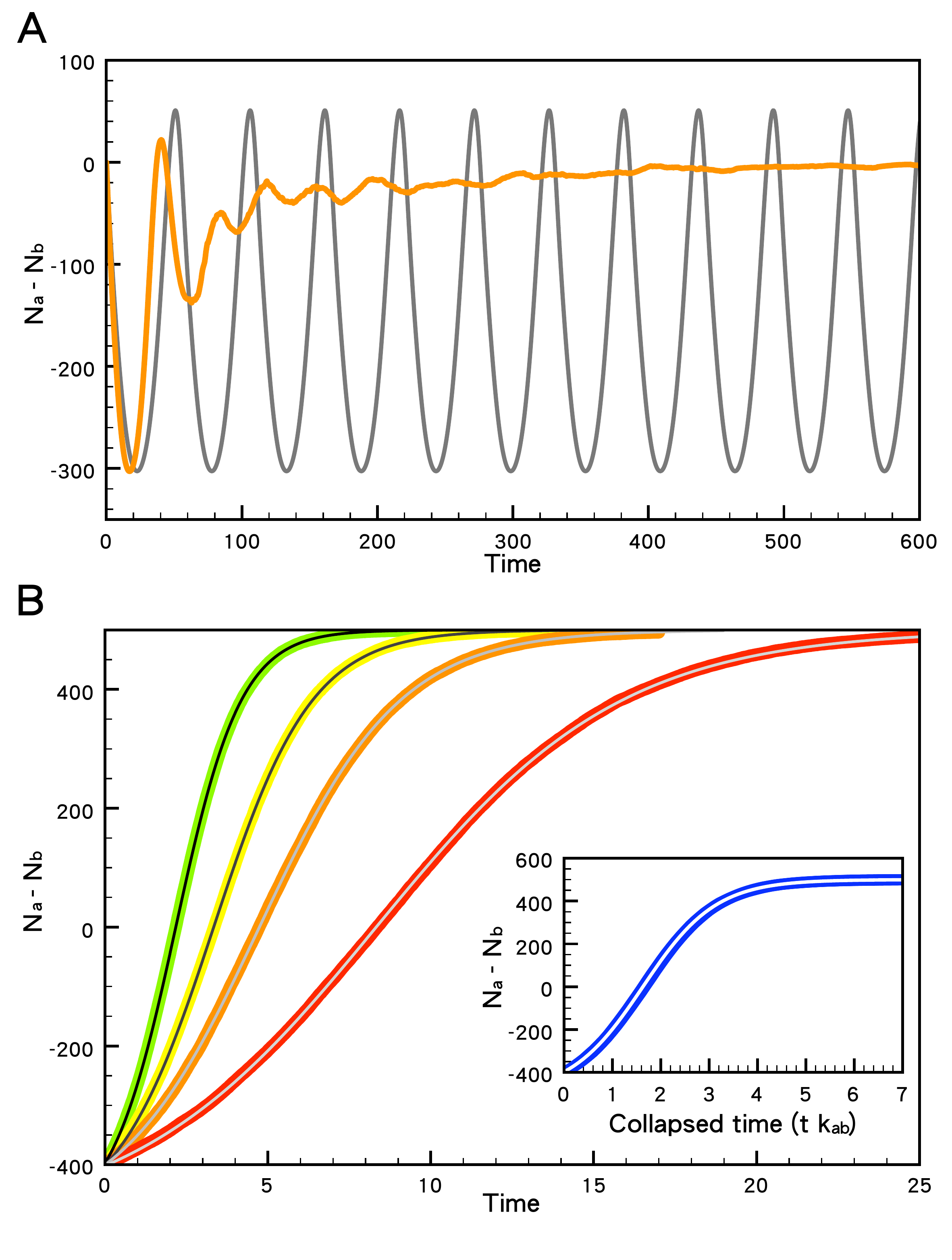}
\caption{
Comparison between the stochastic evolution of $N_a -N_b$ (color solid lines) with finite population size $N=500$ and that from the replicative dynamics (grey lines).
(A) Starting inside the simplex $S_3$, the stable oscillatory solution from the replicator equations is gradually damped out by the intrinsic fluctuations.
(B) On the absorbing boundary, the cyclic competition no longer reigns and the competing mechanism is the usual prey-predator type.
The population evolution in this hierarchical case can be solved analytically for different $k_{ab}=2, 3, 4.5, 8$ (from left to right in grey colors).
The stochastic evolution for the hierarchical competition gives identical results (green, yellow, orange and red curves) to the replicative dynamics. As shown in the inset, all evolution trajectories show nice scaling behavior and can be collapsed onto a universal curve when plotted against the rescaled time $k_{ab}t$.}
\end{figure}

To visualize the origin of the intrinsic fluctuations, it is convenient to plot the evolution trajectory in an appropriate phase space.
Let us ignore the noise terms in the replicator equations momentarily and study the symmetric case with $k_{ab}=k_{bc}=k_{ca}=1/3$.
It is straightforward to solve the replicator equations and obtain the oscillatory solutions for the population ratios as shown in Fig. 1A.
Due to the constraint $a+b+c=1$, the allowed phase space for the evolution trajectories comprises a standard triangle\cite{Nowak06}, known as the simplex $S_3$ with three absorbing corners where only one species survives.
In addition, along the three boundaries where one competing species becomes extinct, the evolution dynamics is just the usual prey-predator type and not cyclic in nature anymore.
The oscillatory solution for the replicator equations traces out a closed contour in Fig. 1B clockwise inside the simplex $S_3$.
When the size of population is finite, the changes of population ratios occur in steps of stride length $1/N$ and form a microscopic triangular lattice inside the simplex.
For visual clarity, the underlying triangular lattice for a small population $N=9$ is plotted in Fig. 1B.
The stochastic processes lead to zigzag hopping on the triangular lattice, while the replicator dynamics predicts a smooth contour inside the simplex.
The deviations between the zigzag hopping and the smooth trajectory is the origin of the intrinsic noises.
It shall be clear that the strength of these noises is of order $1/N$ and will only vanish in the limit of infinite population.

\begin{figure}
\centering
\includegraphics[width=10cm]{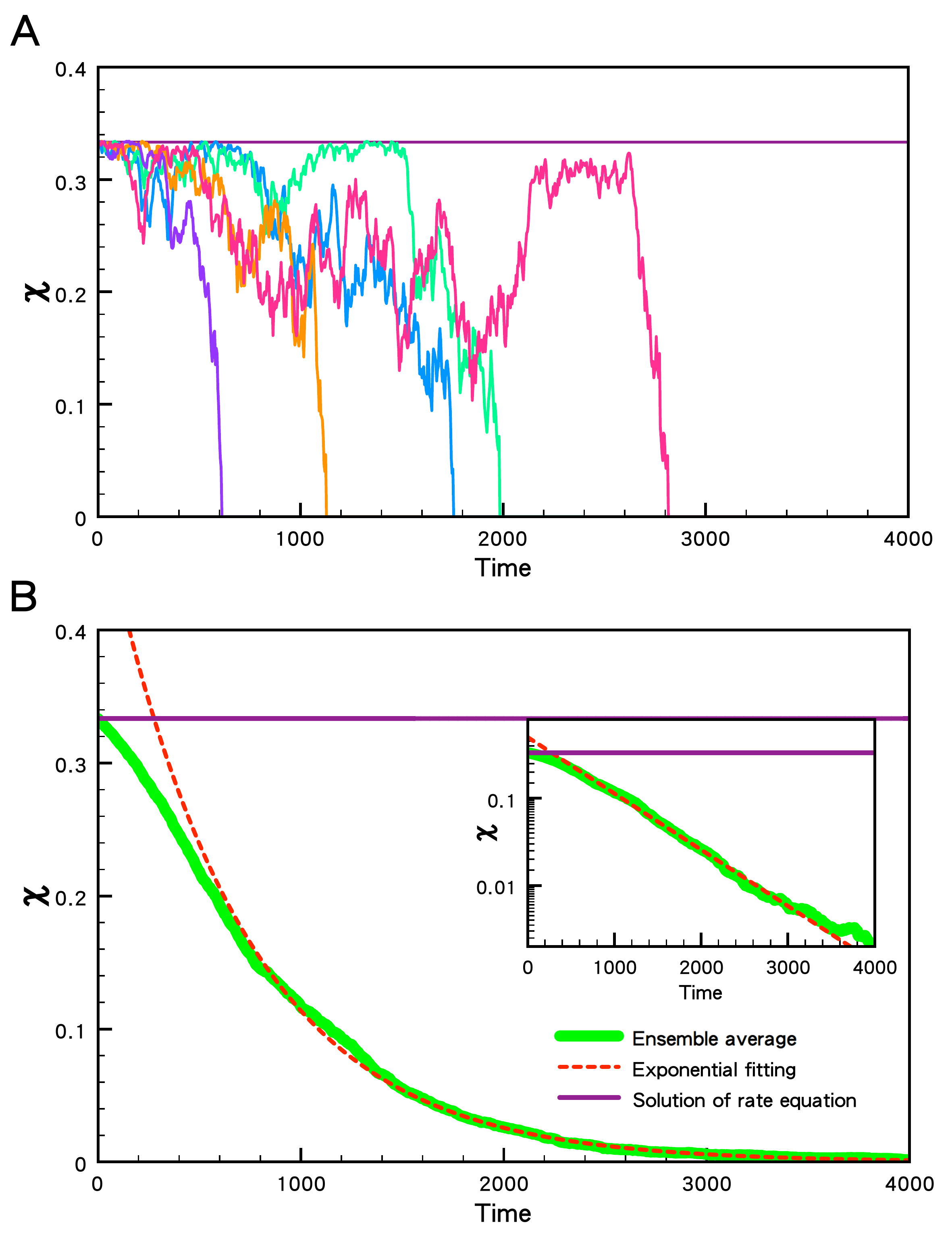}
\caption{The biodiversity indicator $\chi(t)$ of the cyclic-competing ecosystem.
(a) Several single-round evolutions in the numerical simulations are shown in different colors, where ($k_{ab},k_{bc},k_{ca}$)=(0.30,0.35,0.35) and $N =900$ with equal initial subpopulations.
The biodiversity indicator is constant within the replicative dynamics (purple solid line) but driven to extinction $\chi =0$ by random events in the stochastic evolution.
(b) After ensemble averaging through 1000 samples, the biodiversity indicator is found to converge to a smooth exponential decaying function, as corroborated by the straight line on the semi-logarithmic plot in the inset.}
\end{figure}

One may naively expect that the noises make the evolution wandering around the smooth trajectory predicted by the replicative dynamics in a random fashion.
If so, as long as the trajectory is far from the absorbing boundaries of the simplex, the random wandering can be averaged out.
We choose an initial set of the population ratios inside the simplex and solve for the evolution by both stochastic and replicative approaches.
The differences between these two approaches are compared in Fig. 2A.
The replicator equations deliver the oscillatory solutions for the population ratios as expected.
In contrast, the stochastic processes seem to damp out the indefinite oscillations fairly quickly and lead to extinction with only one surviving species. The drastic difference shows that the shot noises provides more than mere random drags to the smooth contour, as is predicted by the replicative dynamics and appropriate when the initial set happens to fall on the absorbing boundaries.

To capture the non-trivial effects of the intrinsic noises, we introduce a biodiversity indicator
\begin{eqnarray}
\chi(t) = a(t)^{k_{bc}} b(t)^{k_{ca}} c(t)^{k_{ab}},
\end{eqnarray}
which is constant within the replicative dynamics and serves as a good quantitative measure for how the biodiversity enervates during stochastic evolution.
Following the standard derivations from the fluctuation-dissipation theorem\cite{Reif08}, we arrive at the central result of the paper,
\begin{eqnarray}
\frac{d \langle \chi \rangle}{dt} \equiv \frac{\langle \chi(t+\tau) \rangle - \langle \chi(t) \rangle}{\tau} = - \frac{1}{N} D(a,b,c) \beta(\chi),
\end{eqnarray}
where the detail steps can be found in SOM.
The factor $D(a,b,c)$ comes from the noise correlator for $\chi$ and is thus positive at all times.
The $1/N$ factor is pulled out to emphasize that the strength of the noise correlator is inversely proportional to the total population.
The key player for driving the dissipative dynamics is the asymmetry in the phase space,
\begin{eqnarray}
\beta(\chi) \equiv - \frac{d \ln L(\chi)}{d\chi} = -\frac{1}{L} \frac{dL}{d\chi},
\end{eqnarray}
where $L(\chi)$ denotes the contour length for the particular $\chi$.
Note that $\chi$ reaches its maximum at the fixed point $(a,b,c) = (k_{bc}, k_{ca}, k_{ab})$ and decreases monotonically when approaching the absorbing boundaries.
It is clear that $L=0$ at the fixed point and increases monotonically to $L=3$ when approaching the boundaries.
As a result, $dL/d\chi <0$, implying $\beta(\chi)$ is positive-definite.
Basically, the coarse-grained fluctuations generate a dissipative drive toward the absorbing boundaries where the phase space is largest.
It is rather remarkable that the direction of dissipative evolution is dictated by the geometric structure of the simplex $S_3$.
When one species is removed from the competition, the phase space is reduced to one of the absorbing boundaries.
The same reasoning can be applied to the hierarchical competition between the remaining two species and predicts no preferential drift direction since the one-dimensional phase space is symmetrical.
The absence of dissipative drive on the absorbing boundaries agrees with the
numerical simulations to be explained later.

\begin{figure}
\centering
\includegraphics[width=10cm]{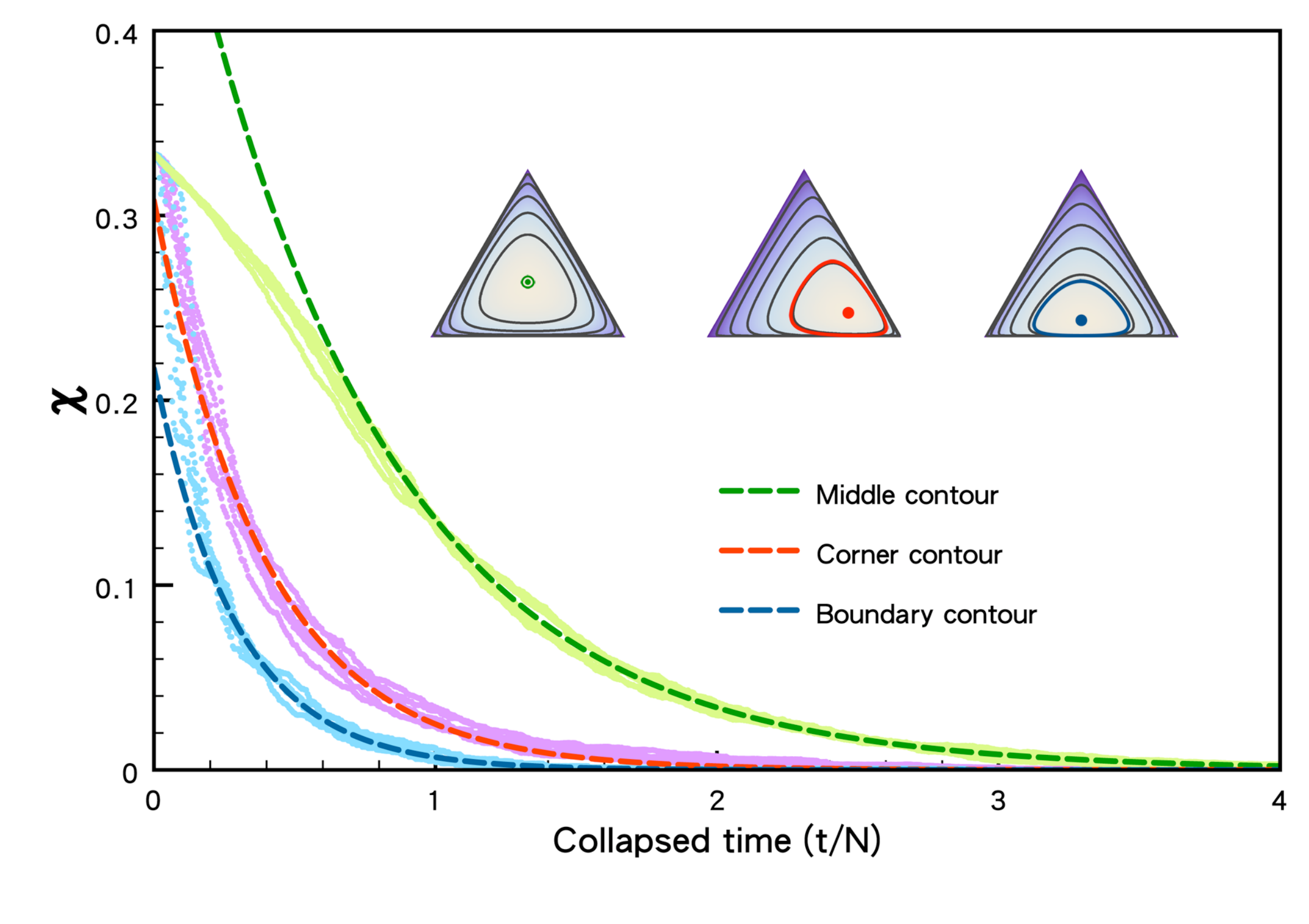}
\caption{The time evolution for the biodiversity indicator $\chi$ with different competing parameters and population sizes. Here we choose three sets of $(k_{ab},k_{bc},k_{ca}) =(0.30,0.35,0.35), (0.15,0.20,0.65)$ and $(0.10,0.45,0.45)$, corresponding to the middle (green), corner (red) and boundary (blue) contours. The population sizes are $N=90, 180, 360, 540, 720, 900$ with equal initial subpopulations. It is rather impressive that the biodiversity indicator always decays exponentially. Furthermore, the scaling behavior for different population sizes is manifest when plotted versus the rescaled time $t/N$. All curves with the same competing parameters fall onto the universal curves, captured well by the exponential decays (dashed lines).}
\end{figure}

Numerical simulations are performed to back up our predictions.
The ensemble-averaged $\chi(t)$ in Fig. 3 indeed reveals the monotonic decreasing trend.
Except for the initial transient period, the dissipative dynamics is well captured by the exponential decay.
This universal exponential decay is due to the simplification of the dissipative equation around small $\chi(t)$.
When close to the absorbing boundaries, the product $D(\chi) \beta(\chi)$ is small and mainly depends on $\chi$.
We can then expand the equation near $\chi =0$ to the linear order,
\begin{eqnarray*}
\frac{d \langle \chi \rangle}{dt} = - \left(\frac{1}{N \tau_{ex}}\right) \chi,
\end{eqnarray*}
where $1/\tau_{ex} = d(D\beta)/d\chi|_{\chi=0}$.
As a result, the extinction process is exponential in the long-time limit and the extinction time $T_{ex} = N \tau_{ex}$ of the ecosystem scales linearly with the system size $N$. We perform extensive numerical simulations in the parameter space and find the exponential form robust.
It is remarkable that the time evolution of the biodiversity indicator shows interesting scaling behaviors with respect to the population size as indicated in Fig. 4.
For a chosen set of competing parameters $(k_{ab}, k_{bc}, k_{ca})$, the biodiversity indicators for different population sizes $N$ can be collapsed onto a universal curve when plotted versus the rescaled time $t/N$. 
Note that the rescaled extinction time $\tau_{ex} = T_{ex}/N$ is closely related the location of the evolutionary trajectory.
As shown  in Fig. 4, for contours deep inside the simplex, the dissipation takes a longer time to drain up the biodiversity.
On the other hand, the dissipation is understandably much quicker when contours are close to the boundaries or corners.
However, despite of the differences in details, the almost-perfect collapse shows the universal exponential decay with $1/N$ scaling.

\begin{widetext}
\begin{table}
\begin{ruledtabular}
\begin{tabular}{l|cccc}
shared properties & cyclic competition & damped SHM\\ \hline\hline
conserved quantity & biodiversity indicator $\chi$ & mechanical energy $E$\\\hline
nature of fluctuations & intrinsic & extrinsic\\ 
 & discreteness of populations & random collisions from reservoir\\\hline
resulting dissipation & ${d\chi\over dt}<0$ & ${dE\over dt}<0$\\ \hline
eventual outcome & definite extinction & come to a stop \\
\end{tabular}
\caption{The close analogy between the cyclic-competing ecosystem and the damped simple harmonic motion (SHM) in a viscous liquid.}
\end{ruledtabular}
\end{table}
\end{widetext}

The macroscopic dissipation emerges from the microscopic fluctuations is best understood in the context of fluctuation-dissipation theorem\cite{Reif08} for general dynamical systems.
Table I provides a clear similarity of the shared properties between the cyclic competition and the damped simple harmonic motion (SHM). However, there are still marked differences which make the dissipative dynamics in the finite-population ecosystem unique.
 In ordinary dynamical systems, the liquid reservoir provides the source of fluctuations and also establishes the asymmetry of the phase space since the degrees of freedom in the reservoir is much larger than those of the studied dynamical system.
However, in the cyclic-competing ecosystem, there is no external reservoir and the fluctuations are intrinsic and originated from the discreteness of species populations.
In addition, the asymmetry of the phase space for the ecosystem is not always guaranteed because there is no external reservoir. 
By now, we have established a new source of dissipations to the non-transitive competitions inside the simplex $S_3$. Although these dissipations are absent for the hierarchical competitions on the absorbing boundaries, they may be resurrected when the structure of the phase space is changed upon the inclusion of other competing mechanisms.
An example of this is the Lotka-Volterra model in which both selection and reproduction of preys and predators are taken into account.
The evolution trajectory becomes two dimensional and the phase space shares the same asymmetrical structure as for the cyclic-competing ecosystems. 
We thus expect similar dissipative dynamics will occur and bring down the biodiversity.
It is rather profound that the asymmetrical structure of the phase space pins down the direction for the dissipative evolution toward extinction.

In conclusion, we investigate the intrinsic fluctuations from the discreteness of the populations and show how the dissipative dynamics for the biodiversity indicator emerges.
Our findings pave a different route to address the intrinsic noises beyond the replicator dynamics and deepen our understanding for the stability of biodiversity and the evolutionary dynamics toward extinction in ecosystems with finite populations.

We acknowledge supports from the National Science Council in Taiwan through grants 
95-2112-M007-046-MY3 (YCL and TMH) and NSC-97-2112-M-007-022-MY3 (HHL). Financial supports and friendly environment provided by the National Center for Theoretical Sciences in Taiwan are also greatly appreciated.

\newpage

\noindent \textbf{\Large Supplementary Online Materials}

\section{Intrinsic fluctuations}

We would like to derive the intrinsic noises from the discreteness of the populations. For instance, the population for species $A$ after each stochastic evolution step changes $\Delta N_a = 1, 0, -1$ with probabilities $P_1, P_0, P_{\bar 1}$ respectively. We can separate the changes into the average part and the fluctuations,
\begin{eqnarray}
\Delta N_a(\tau) = R_{ab} - R_{ca} + \xi_a(\tau),
\end{eqnarray}
where $R_{ab} = k_{ab} a b$ and $R_{ca} = k_{ca} c a$. The simulation time $\tau = 1,2,3,...$ denotes the generation of the stochastic evolution. By construction, the probability distribution for the intrinsic noise is
\begin{eqnarray}
&P_1=R_{ab}& \to \quad \xi_a = 1-(R_{ab}-R_{ca}),\\
&P_{\bar 1}=R_{ca}& \to \quad \xi_a = -1-(R_{ab}-R_{ca}),\\
&P_0=1-R_{ab}-R_{ca}& \to \quad \xi_a = 0-(R_{ab}-R_{ca}).
\end{eqnarray}
It is easy to check that the average of the noise is indeed zero. The correlation function for the noise is straightforward to compute,
\begin{eqnarray}
\langle \xi_a(\tau) \xi_a(\tau') \rangle = \delta_{\tau \tau'} \sum_{X=1, {\bar 1}, 0} P(X)\: \xi^2_a(X).
\end{eqnarray}
After some algebra, the noise correlator for species $A$ is
\begin{eqnarray}
\langle \xi_a(\tau) \xi_a(\tau') \rangle = \delta_{\tau \tau'} \bigg[ R_{ab}+ R_{ca}
- (R_{ab}-R_{ca})^2 \bigg]
\end{eqnarray}
Similarly, one can compute the other noise correlation functions for species $B$ and $C$,
\begin{eqnarray}
\langle \xi_b(\tau) \xi_b(\tau') \rangle = \delta_{\tau \tau'} \bigg[ R_{bc}+ R_{ab}
- (R_{bc}-R_{ab})^2 \bigg],
\\
\langle \xi_c(\tau) \xi_c(\tau') \rangle = \delta_{\tau \tau'} \bigg[ R_{ca}+ R_{bc}
- (R_{ca}-R_{bc})^2 \bigg].
\end{eqnarray}
 
Suppose the population size is large enough that the finite difference can be approximated by the continuous differential,
\begin{eqnarray}
\frac{da}{dt} \approx \frac{\Delta a}{\Delta t} = \frac{\Delta N_a}{\Delta \tau} = R_{ab} - R_{ca} + \xi_a(t),
\end{eqnarray}
where $\Delta a = \Delta N_a/N$ and $\Delta t = \Delta \tau /N = 1/N$. Therefore, the replicator equation is recovered with an intrinsic noise $\xi_a$ due to the discreteness of the population. Note that the strength of the intrinsic noises exhibits a $1/N$ factor in the continuous limit,
\begin{eqnarray}
\langle \xi_a(t) \xi_a(t') \rangle = \frac{1}{N} \bigg[ R_{ab}+ R_{ca}
- (R_{ab}-R_{ca})^2 \bigg] \delta(t-t') .
\end{eqnarray}
The replicator equations with intrinsic noises for species $B$ and $C$ can be derived likewise.

\section{Dissipative dynamics}

In the following, we shall borrow the lesson of fluctuation-dissipation theorem in physics\cite{Reif08} to show the emergence of the dissipative dynamics for the biodiversity indicator. Since the biodiversity indicator $\chi(t)$ is a constant of motion for the replicator equations, the dynamics is driven purely by the corresponding noise $\xi_\chi(t)$. Integrating the dynamical equation for $\chi(t)$, the finite difference after ensemble average is
\begin{eqnarray}
\langle \chi(t+\tau) \rangle - \langle \chi(t) \rangle =
\int_t^{t+\tau} dt' \langle \xi_\chi(t') \rangle.
\end{eqnarray}
It is important to emphasize that $\tau$ is much smaller than the macroscopic time scale we are interested in, but much larger than the microscopic time scale for evolution steps,
\begin{eqnarray}
\tau^* \ll \tau \ll \Delta t.
\end{eqnarray}
Since $\tau \gg \tau^*$, the ensemble average makes sense at time $t'$ with a modified probability distribution $P(t')$ which differs from the original one $P(t)=P_0$.

Assuming the probability distribution function $P(t')$ is proportional to the corresponding phase space $L(\chi+\Delta \chi)$,
\begin{eqnarray}
\frac{P(t')}{P(t)} = \frac{L(\chi+\Delta \chi)}{L(\chi)} \approx e^{-\beta \Delta\chi}.
\end{eqnarray}
The finite difference in the biodiversity indicator is
$\Delta \chi = \int_t^{t'} \xi_\chi(t'') dt''$ and
the function $\beta(\chi) \equiv - d(\ln L)/d\chi >0$ because $L(\chi)$ decreases as $\chi$ increases. Putting all pieces together, we obtain the important identity
\begin{eqnarray}
\langle \xi_\chi(t') \rangle = \langle \xi_\chi(t') e^{-\beta \Delta\chi}\rangle_0
\end{eqnarray}
The ensemble average can be computed by expanding the exponential,
\begin{eqnarray}
\langle \xi_\chi(t') \rangle &\approx&
\langle \xi_\chi(t') (1- \beta \Delta \chi) \rangle_0 
= - \beta(\chi) \langle \xi_\chi(t') \Delta \chi \rangle_0
\nonumber\\
&& \hspace{-2cm} = -\beta(\chi) \int_t^{t'} dt'' \langle \xi_\chi(t') \xi_\chi(t'') \rangle_0
= - \frac{1}{N} \beta(\chi) D(a,b,c),
\end{eqnarray}
where both $\beta(\chi)$ and $D(a,b,c)$ are positive definite. Finally, the time evolution of the biodiversity indicator is
\begin{eqnarray}
\frac{d \langle \chi \rangle}{dt} \equiv \frac{\langle \chi(t+\tau) \rangle - \langle \chi(t) \rangle}{\tau} = - \frac{1}{N} \beta(\chi) D(a,b,c) <0.
\end{eqnarray}
It is quite remarkable that we prove that the ecosystem tends to evolute toward extinction after coarse-graining the intrinsic fluctuations from the discreteness of the populations.

\end{document}